\documentstyle[preprint,aps]{revtex}
\def\be{\begin{equation}}
\def\ee{\end{equation}}
\def\ba{\begin{eqnarray}}
\def\ea{\end{eqnarray}}
\def\ep{\epsilon}
\def\nn{\nonumber}
\def\ov{\over}
\def\bb{\mbox{$\displaystyle \{ $ }}
\def\eb{\mbox{$\displaystyle \} $ }}

\begin{document}
\draft
\preprint{IP/BBSR/94-59 }
\title {   The Pressure of Hot $g^2 \Phi^4$ Theory at order $g^5$ }
\author{ Rajesh Parwani and Harvendra Singh}
\address{Institute of Physics, Bhubaneswar-751 005,
INDIA.\footnote{e-mail: parwani@iopb.ernet.in and hsingh@iopb.ernet.in}}
\maketitle
\begin{center}
01 Nov 1994
\end{center}
\begin{abstract}

The order $g^5$ contribution to the pressure of massless $g^2 \phi^4$
theory at nonzero temperature is obtained explicitly. Lower order contributions
are reconsidered and two issues leading to the optimal choice of
rearranged Lagrangian for such calculations are clarified.

\end{abstract}
\vspace{2 cm}
\pacs{PACS NO. 12.38.Mh, 11.10.Jj, 12.38.Cy}
\narrowtext

\newpage

\section {Introduction}
\label{int}

\par Rapid progress has been made recently in computations of the
free energy density, at nonzero temperature $(T)$, of massless $g^2
\phi^4$ theory \cite{FST}, quantum electrodynamics (QED) \cite{CP} and
quantum chromodynamics (QCD) \cite{AZ}, to three-loop (fourth order in
coupling). For QED, the fifth $(e^5)$ order contribution has also
been obtained \cite{P1} by dressing the photon lines of the three-loop
diagrams.

Compared to QED, a fifth order calculation in QCD will be more
involved because gluonic self-interactions imply that many more lines
in any three-loop diagram can be soft (i.e. at zero Matsubara
frequency in the imaginary time formalism), and so must be dressed in
order to obtain the full $g^5$ contribution. In this respect a fifth
order calculation in QCD will resemble the same order calculation in
$g^2 \phi^4$ theory. In this paper we will compute the order
$g^5\,T^4$ contribution to the pressure of massless $g^2 \phi^4$
theory and show that by an optimal choice of rearranged Lagrangian it
is possible to obtain analytical results without too much effort.

The theory we are concerned with is defined by the Euclidean Lagrangian

\be
{\cal{L}}_0= {1 \over 2} (\partial\phi)^2 + {g^2 \mu^{2 \ep}\over 4!}
\phi^4\,\,,
\label{1.1}
\ee
where $\mu$ is the mass scale of dimensional regularisation($d = 4 -
2\epsilon$)
. We use the imaginary time formalism in which the energies take
discrete values, $2\pi n T$, $n \in \cal{Z}$. For  perturbative
calculations beyond leading order, it is necessary to take into
account in a systematic manner the non-negligible collective
effects \cite{BP}. For the theory defined by eq.(\ref{1.1}) this means
that one uses instead the shifted Lagrangian \cite{P2}

\be
{\cal{L}}^{\prime}_0 = \left({\cal{ L}}_0 + {1\over2} m^2 \phi^2\right)
-{1\over2} m^2 \phi^2
\label{1.2}
\ee
with $m^2=g^2 \, T^2/ 24 $ the thermal mass generated at one-loop. The
term within brackets in eq.(\ref{1.2}) defines a dressed
propagator  while the last term is a new two-point vertex which
prevents overcounting. With eq.(\ref{1.2}) one can proceed to
calculate any Green's function in the theory, order by order, in a
consistent way. However for the calculation of Green's functions with
{\it{static}} (zero energy) external legs, eq.(\ref{1.2}) is not very
economical since it involves some extraneous resummation.

Recall (see, for example, \cite{CP,AZ}) that for a static Green's function, its
physical definition is
already given in imaginary time, without the need to analytically
continue to real time. Thus the power counting of infrared (IR)
divergences may be safely done using the Euclidean propagators with
discrete energies. Then, only the propagators at zero Matsubara
frequency do not have an IR cut-off of order $T$,
and it is only for these zero modes that the thermal mass $\sim g\,T$
is a relevant infrared cut-off. Thus instead of eq.(\ref{1.2}) one
can use

\be
{\cal{L}} = \left({\cal{L}}_0 + {1\over2} m^2 \phi^2 \delta_{p_{0},0} \right)
- {1\over2} m^2 \phi^2 \delta_{p_{0},0}
\label{1.3}
\ee
for {\it{static}} calculations, with $p_{0}$ the energy in
Fourier space of $\phi(x)$. The Lagrangian in eq.(\ref{1.3}) is
precisely what is suggested by the Braaten-Pisarski \cite{BP}
resummation scheme whiu
 involves dressing 'soft' lines with
'hard-thermal loops'. For the calculation of {\it{static}}
quantities in {\it imaginary time}, the only soft-line
is the zero-mode
propagator and the only hard-thermal loop is the static one-loop
thermal mass.

We remark that a resummation as in eq.(\ref{1.3}), which involves
dressing only the zero-modes, was used very effectively by Arnold and
co-workers \cite{AE,AZ} for their free energy calculations in gauge theories,
but they used
the more general expression (\ref{1.2}) (as in Ref.\cite{FST}) for their
scalar free-energy calculation. The advantage of using minimal
resummation (\ref{1.3}) also for the scalar case can be seen by
the following example. The propagator from eq.(\ref{1.3}) is

\be
\Delta(K)\equiv { (1 - \delta_{k_{0},0})\over K^2} +
{\delta_{k_{0},0}\over k^2 + m^2}\,\,.
\label{1.4}
\ee
Then the one-loop integral

\ba
\int[dK] \Delta(K) &=& \int[dK] {(1 - \delta_{k_{0},0})\over K^2} +
T\int(dk){1\over k^2 + m^2} \label{1.5} \\
&=& {1 \ov 2} \pi^{{d-5 \ov 2}} T^{d-2} \zeta(3-d) \ \Gamma\left({3-d
\ov 2}\right) \, + \, { T \ov 4 \pi} \left(m^2 \over 4\pi\right)^{{d-3
\ov 2}} \ \Gamma\left( {3-d \ov 2} \right) \; ,
\ea
where we have used the notation
\be
\int[dK]\equiv T \sum_{k_0 (\bf{even})} \int (dk) \equiv T \sum_{k_0
(\bf{even})} \int { d^{d-1} k \over (2 \pi)^{d-1} }\,\,\,,
\nonumber
\ee
with $K_{\mu} = (k_{0},\vec{k})$. Notice that the integrals in
eq.(\ref{1.5}) were computable in closed form and the result is a term
of order $g^0$ and another of order $g$. Since the bosonic one-loop
integral (\ref{1.5}) and others related to it occur frequently
in Figs.1(a-g) , one easily sees that all those diagrams may be
evaluated easily. By contrast the propagator of eq.(\ref{1.2})
is $1/ {(K^2 + m^2)}$ and its one-loop integral is only available
as a high temperature expansion \cite{AZ}, making the evaluation of
the diagrams of theory (\ref{1.2}) more involved.

Another point that needs clarifying is the choice of $m^2$ in
eq.(\ref{1.3}). In four dimensions, $m^2 = {g^2 T^2 \over 24}$.
However since we are using dimensional regularisation, one might
wonder if the value of $m^2$ in $d$-dimensions should be used. For
example, in \cite{FST} the value $\overline{m^2} = {g^2
T^{2-2\epsilon}\over 24} \mu^{2\epsilon}$ was used (it being the hard
thermal loop in $4 - 2 \epsilon$ dimensions), while in \cite{AZ} the
full one-loop value in $d$-dimensions was used. Will these different
choices affect the final result? If one is interested in renormalised
values as $\epsilon \to 0$, the answer is no. Here is the proof: Let
the resummation in eq.(\ref{1.3}) be done using some $m^2 (\epsilon)$
with $m^2 (0)= {g^2 T^2 \over 24}$. Now keep the $\epsilon$ - dependence
of $m^2 (\epsilon)$ implicit, even when it hits $1/ \ep$
terms. Since the full Lagrangian in eq.(\ref{1.3}) is massless, no
ultra-violet mass renormalisation is needed in perturbative
calculations using dimensional regularisation. Therefore any
$m^2(\epsilon)\over \epsilon$ terms generated when calculating a
renormalised quantity (i.e. after including coupling-constant renormalisation)
must mutually cancel. Finally only terms of the form
$m^2(\epsilon) \epsilon^n$ $(n \geq 0)$ will appear and then, as
$\epsilon \to 0$, only $m^2(0)$ survives. End of proof.

In explicit calculations one finds that if the $\epsilon$-dependence
of $m^2(\ep)$ is expanded  out then extra finite ($\ep^0$) terms are
generated from some diagrams because of the $\ep$-dependence of the
mass in the propagator, but these will cancel , order by order in
$g$, with similar terms generated from $\ep$-dependence of the
thermal counterterm. The fact that one can use $m^2(0)$ rather than
some $m^2(\ep)$ clearly simplifies calculations and avoids redundant
cancellations. On the other hand one can exploit the proof in the
last paragraph to provide a cross-check on calculations. That is, one
can use $\hat{m}^2 = {g^2 T^2 \over 24} (1 + \ep A)$, where
$A$ is an {\it{arbitrary}} regular function of $\ep$ ( which may also
depend on $g$ and $T/\mu$), in eq.(\ref{1.3}) and verify that one's
final renormalised result is independent of $A$ as $\ep \to 0$. Such a
check had been performed \cite{P3} for the pole of the propagator of
$g^2 \phi^4$ theory to two-loop \cite{P2}, and we have also done it
for the calculations in this paper. However in order to emphasize the
simplicity of using $m^2(0)$, we will  present here the results for
this case only.

In the next section we reconsider the calculations of the pressure to
order $g^4$ using eq.(\ref{1.3}) and recover the results of
\cite{FST,AZ}. Then in section-III the $g^5$ terms are obtained and
the results are summarised in section-IV. We conclude with some
comments in section-V.

\section {Lower Orders}
\label{LO}

\par The diagrams which contribute to the pressure to order $g^5$ are
shown in Fig.($1$). The $g^4$ terms from these diagrams were
extracted in refs. \cite{FST,AZ}. Here we reconsider those $g^4$
terms using the minimal Lagrangian, eq.(\ref{1.3}), with $m^2 = {g^2
T^2 \over 24 }$. The reader comparing the results here with those of
\cite{FST,AZ}
should note this difference; order by order (rather than diagram by
diagram) our results here agree with those of \cite{FST,AZ}.
Diagram($a$) contributes

\ba
{\cal{P}}_a &=&  -{1 \over 2} \int[dP] \ln \Delta^{-1} (P) \nonumber\\
&=& -{1 \over 2} \int [dP] \ln \{ P^2 (1 - \delta_{p_0,0}) +
(P^2 + m^2) \delta_{p_0,0}\} \nonumber \\
&=& -{1\over2}\int[dP] \ln P^2 - {1\over2}\int[dP] \ln\left(1+ {m^2
\delta_{p_0,0}\over
P^2 }\right) \nonumber \\
&=& -{1\over2}\int[dP] \ln P^2 - {T\over2}\int(dp) \ln \left( 1 +
{m^2\over p^2} \right) \label{2.1} \\
&=& T^d \pi^{-d/2} \xi (d) \Gamma (d/2) - {T\over 2} \Gamma \left( {d-1
\over 2}\right) \left({m^2\over 4\pi}\right)^{(d-1)/2} \nonumber \\
&=&{\pi^2 T^4 \over 90} + {\pi^2 T^4 \over 9 \sqrt{6}} \left({g\over
4 \pi}\right)^3 + O(\ep) \label{2.2}
\ea
In eqs.(\ref{2.1},\ref{2.2}), the first term represents the ideal gas
contribution while second term is the plasmon contribution (in
dimensional regularisation, see \cite{CP}) obtained by dressing the
zero-mode of the one-loop diagram.

The contributions of diagrams (b) through (g) are:

\ba
\mu^{2\ep}({\cal{P}}_b + {\cal{P}}_c)&=& -{g^2\over 8}
\mu^{4\ep} \left[\int[dP] \Delta(P)\right]^2 + {m^2\over 2} \mu^{2 \ep}\int[dP]
\Delta(P) \delta_{p_0,0}\nn\\
&=& -{g^2 T^4 \over 2^7 \ 3^2} - {g^4 T^4 \over 2^{10}\ 3\ \pi^2} + O(\ep)
\label{2.3}\\
& &\nn\\
\mu^{2\ep}{\cal{P}}_d &=& -{\mu^{2\ep}\over 8}\left(3g^4\over 32
\pi^2 \ep \right)\left[\int[dP] \Delta(P)\right]^2 \nn\\
&=& {1\over\ep}\,{g^4 T^4 \over 2^{12}\ \pi^2}
\left( -{1\over3}+{g\over\pi\sqrt{6}} - {g^2 \over 2^3 \pi^2} \right)\nn\\
 && - {g^4 T^4 \over {3 \ \pi^2 \ 2^{12}}}
\left( 4\ln {\mu\ov T} + 4 - 2\gamma -2\ln 4\pi +
4{\zeta^{\prime}(-1)\over\zeta(-1)} \right) \nn\\
 && +{g^5 T^4\over 2^{12}\ \pi^3 \ \sqrt{6}} \left( 4\ln {\mu\over T} +
4- 2\gamma +2{\zeta^{\prime}(-1)\over\zeta(-1)} + \ln {6\over g^2}
\right) + O(g^6) \ . \label{2.4}\\
 && \nn\\
\mu^{2\ep}({\cal{P}}_e + {\cal{P}}_f + {\cal{P}}_g) &=& {\mu^{2\ep}
\over 4 } \int[dP] \delta_{p_0,0} \Delta^2 (P)\left[ m^2 - {g^2
\mu^{2\ep}\over2 }  \int[dK]\Delta(K)\right]^2 \nn \\
&& \ + { \mu^{2\ep}\over 4} \int[dP] (1-\delta_{p_0,0})\Delta^2(P)
\left[{g^2\mu^{2\ep} \over2}\int[dK] \Delta(K) \right]^2 \nn\\
&=& {1\over\ep}{g^4 T^4\over 3\ 2^{12}\ \pi^2}\left({1\over3} -
{g\over\pi\sqrt{6}} + {g^2\over 2^3 \pi^2}\right) \nn\\
& &+{g^4 T^4 \over 3^2\ \pi^2\ 2^{12}} \left(6\ln{\mu\ov T} +4-\gamma
-3\ln4\pi +4{\zeta^{\prime}(-1)\over\zeta(-1)}\right) \nn\\
& & -{g^5 T^4 \over3\ 2^{12}\ \pi^3\ \sqrt{6}} \left( 6\ln{\mu\ov T} +1
-\gamma +\ln{3\over2 \pi g^2} +
2{\zeta^{\prime}(-1)\over\zeta(-1)}\right)+ O(g^6). \label{2.5}
\ea
Note that in ${\cal{P}}_d$ only the {\it{one}}-loop ultra-violet (UV)
coupling constant renormalisation counterterm is used. (We are
also using minimal subtraction). The
$g^4/\ep$ piece is required to cancel similar pieces from diagram (g)
and (h). The $g^5/\ep$ and $g^6/\ep$ divergences in eq.(\ref{2.4})
are due to the mixing of IR resummation effect with the UV
renormalisation and will  cancel against similar
terms generated from diagrams (g) and (h).

The order  $g^4$ contribution from diagram (h) is obtained by
setting $m = 0$ in the propagators since the integrals are IR finite.
The result ${\cal{P}}_{h4}$ has been
evaluated analytically by Arnold and Zhai and we simply quote their
value (which agrees with earlier semi-analytical  evaluations
in \cite{FST,CP}),

\be
\mu^{2\ep} {\cal{P}}_{h4}={g^4 T^4 \over 3^3\ 2^{12}\
\pi^2}\left({6\over\ep} +18\ln {\mu^2\over 4\pi T^2}
-12{\zeta^{\prime}(-3)\over\zeta(-3)} +
48{\zeta^{\prime}(-1)\over\zeta(-1)} -18\gamma +{182\over5}\right) \,\, .
\label{2.6}
\ee
Thus the sum of diagrams up to order $g^4$ is

\ba
{\cal{P}}_4 &=& {\pi^2 T^4 \over 9}\bb {1\over10} -
{1\over8} \left(g\ov 4\pi\right)^2 + {1\ov \sqrt{6}}\left(g\ov
4\pi\right)^3 \nn\\
& & - \left(g\ov 4\pi\right)^4 \left[ -{3\ov 16}\ln{\mu^2\ov
4\pi T^2} + {1\ov 4}{\zeta^{\prime}(-3)\over\zeta(-3)} -{1\ov
2}{\zeta^{\prime}(-1)\over\zeta(-1)}+ {\gamma\ov 16} + {59\ov
120}\right] \eb \nn\\
& &+ O \left({g^5\ov \ep}\,,g^5\,,{g^6\ov \ep}\right)  \,\, .
\label{2.7}
\ea

The terms to fourth order agree with \cite{FST,AZ}.

\section{Fifth Order}
\label{FFT}

We now pick up the subleading pieces from diagram (h). For this we
first rewrite $\Delta(K)$, given in eq.(\ref{1.4}), as (c.f.
Ref.\cite{P1})

\ba
\Delta(P)&=& {1\over P^2} - {m^2 \ \delta_{p_0,0}  \over p^2(p^2 +
m^2)} \nonumber \\
&\equiv&\Delta_0(P) + \Delta^{\ast}(P) \,\, . \label{3.1}
\ea
Then

\ba
\mu^{2\ep} {\cal{P}}_h &=& {g^4\mu^{6\ep}\over 48} \int [dK dQ dP]
\Delta(K) \Delta(Q) \Delta(P) \Delta(K+Q+P) \nn\\
&\equiv&{1\over48}\{ I_0 + 4 I_1 + 6 I_2 + 4 I_3 + I_4 \} \,\, , \label{3.2}
\ea
where
\ba
I_0&=& g^4 \mu^{6\ep} \int[dK\,dQ\,dP] \Delta_0(K) \Delta_0(Q)
\Delta_0(P) \Delta_0(K+Q+P) \,\, ,\nn\\
I_1&=& g^4 \mu^{6\ep} \int[dK\,dQ\,dP] \Delta^\ast(K) \Delta_0(Q)
\Delta_0(P) \Delta_0(K+Q+P) \,\, ,\nn\\
I_2&=& g^4 \mu^{6\ep} \int[dK\,dQ\,dP] \Delta^\ast(K) \Delta^\ast(Q)
\Delta_0(P) \Delta_0(K+Q+P) \,\, ,\nn\\
I_3&=& g^4 \mu^{6\ep} \int[dK\,dQ\,dP] \Delta^\ast(K) \Delta^\ast(Q)
\Delta^\ast(P) \Delta_0(K+Q+P) \,\, ,\nn\\
I_4&=& g^4 \mu^{6\ep} \int[dK\,dQ\,dP] \Delta^\ast(K) \Delta^\ast(Q)
\Delta^\ast(P) \Delta^\ast(K+Q+P) \,\, .\nn
\ea
The integral $I_0$ contributes to ${\cal{P}}_{h4}$ and was considered in the
last
section. We will now extract the order $g^5$, $g^5/\ep$ and
$g^6/\ep$ pieces from $I_1$ through $I_4$. Though our final
objective is  to calculate the pressure only to fifth order, we have
to ensure that all  subleading divergences  such as $g^6/\ep$
(see eq.(\ref{2.4})) cancel (there
are no  divergences beyond $g^6/\ep$ from these diagrams).

Consider
\be
I_1=-g^4 \mu^{6\ep} m^2 T \int {(dk)\over k^2(k^2 + m^2)} \int [dQ \,
dP] {\delta_{k_0,0}\over Q^2 P^2 (K + P + Q)^2} \,\, .
\label{i1}
\ee
Scaling $ \vec{k}\to m\vec{k}$ gives,
\be
I_1 = - g^4 \mu^{6\ep} m^{1-2\ep} T \int {(dk)\over k^2 (k^2+1)} \int
[dQ\,dP] {1\over {Q^2 P^2 \left[(q_0 + p_0)^2 +
(\vec{q} + \vec{p} + m\vec{k})^2 \right]}} \; .
\label{3.3}
\ee

Since the external coefficient is $O(g^5)$, we need only the order
$g$ piece from the $(Q,P)$ integrals. Write the $(Q,P)$ integrals as
(the following  discussion parallels that of $I_{sun}$ in \cite{AZ})
\be
T^2\int {(dq \, dp) \over {q^2 p^2 (\vec{q}+\vec{p}+\vec{mk})^2}} + \int
[dQ\,dP] {
(1-\delta_{q_0,0}\,\delta_{p_0,0}) \over {Q^2 P^2 \left[ (q_0 + p_0)^2 +
(\vec{q} + \vec{p} + m\vec{k})^2\right]}} \; .
\label{3.4}
\ee
As the  second term above is IR safe, one can expand the
denominator
\ba
{1\over (q_0 + p_0)^2 +
(\vec{q}+\vec{p}+\vec{mk})^2}&=&{1\over(Q+P)^2}\left[1+{m^2 k^2 +
2 m \vec{k}\cdot(\vec{q} + \vec{p}) \over (Q+P)^2} \right]^{-1} \nn \\
&=& {1\over (Q+P)^2} \left[ 1 - {2 m \vec{k}\cdot(\vec{q} + \vec{p}) \over
(Q+P)^2} +
O(m^2)\right] \; , \nn\\
\ea
so that the second integral in eq.(\ref{3.4}) is
\be
\int[dQ\,dP] {1 - \delta_{q_0,0}\,\delta_{p_0,0} \over {Q^2 P^2 (Q+P)^2}} - 2m
\int
[dQ\,dP] {1 - \delta_{q_0,0}\,\delta_{p_0,0} \over {Q^2 P^2 (Q+P)^4}}
\vec{k}\cdot(\vec{q} + \vec{p}) +
O(m^2) \, .
\label{3.5}
\ee
The first term in eq.(\ref{3.5}) vanishes in DR \cite{AE,CP,AZ}
while the second term
vanishes when the final $\vec{k}$ integrals are performed in
eq.(\ref{3.3}). Hence
\be
I_1 = -g^4 \mu^{6\ep} m^{1-2\ep} T^3 \int {(dk)\over k^2(k^2 +1)} \int
{(dp \,dq) \over p^2 q^2 (\vec{q} + \vec{p} + m\vec{k})^2} + O(g^7)
\; . \label{3.5b}
\ee
The $(p,q)$ integrals are logarithmically sensitive to $m$
in the infrared. They also have a logarithmic UV singularity in $d-1 =3$
dimensions and so must be evaluated in $d=4-2\ep$ dimensions.
After a standard  evaluation of the $(q,p)$ integrals
one can perform the final $\vec{k}$ integrals easily by keeping
only the terms to $O(\ep^0)$. We obtain

\ba
I_1 &=& - {g^5 T^4 \over \sqrt{24}(32\pi^2)(8\pi)} \left(\mu\over m
\right)^{6\ep} \left[
{1\over\ep} + (8 - 3\gamma + 4\ln2 + 3\ln\pi) + O(\ep)\right] + O(g^7)\,\,.
\label{3.6}
\ea
\newpage
Next consider $I_2$:
\ba
I_2&=& g^4 m^4 \mu^{6\ep} T^2 \int {(dk\,dq)\over k^2 q^2
(k^2+m^2)(q^2+m^2)} \int{[dP] \delta_{k_0,0}\,\delta_{q_0,0}\over
P^2 (K+Q+P)^2} \nn\\
&=& g^4 m^{2(1-2\ep)} \mu^{6\ep} T^2 \int {(dk\,dq) \over k^2 q^2
(k^2+1)(q^2+1)} \int{[dP]\over P^2} {1\over p_0^2 +
(\vec{p}+m\vec{k}+m\vec{q})^2} \; .\nn
\ea
The $P$-integral is
\ba
& &T\int(dp){1\over p^2(\vec{p}+m\vec{k}+m\vec{q})^2} +
\int{[dP] \over P^2}
{(1-\delta_{p_0,0})\over p_0^2 + (\vec{p}+m\vec{k}+m\vec{q})^2} \nn\\
&=& m^{d-5} T \int (dp)
{1\over p^2 (p+k+q)^2} + \int[dP] {1 - \delta_{p_0,0} \over P^4} + O(m)\,\,.
\nn
\ea
Thus
\ba
I_2&=& \mu^{6\ep} g^4 m^{2(1-2\ep)} T^2 \int (dk\,dq){1\over k^2 q^2
(k^2+1)(q^2+1)} \int[dP] {1 - \delta_{p_0,0} \over P^4} \nn\\
& &+ \ \mu^{6\ep} g^4 m^{1-6\ep} T^3 \int {(dk\,dp\,dq)\over k^2 p^2 q^2
(k^2+1)(p^2+1) (\vec{k}+\vec{p}+\vec{q})^2} + O(g^7) \, .
\label{3.7}
\ea
The first line in (\ref{3.7})  is of  order $O(g^6)$. Since we  require
at most the $O(g^6 / \ep)$ piece, we can set $d = 4$ everywhere there
except  in the $P$-integral which gives the pole
\be
\int[dP]{1 - \delta_{p_0,0} \over P^4}= {\pi^2\over
(2\pi)^4}{T^{-2\ep} \over\ep} + O(\ep^0) \; ,
\nn
\ee
and so  the first line of eq.(\ref{3.7}) is

\be
{g^4 m^2 T^2 \over 2^8\ \pi^4\ \ep} + O(g^6 \ep^0) \; .
\label{3.8}
\ee
The integral in the second line of eq.(\ref{3.7}) is finite as $\ep
\to 0$. Therefore  we need to evaluate
\be
{1\over(2\pi)^9} \int {d^3 k \,d^3p \,d^3q \over
k^2(k^2+1)p^2(p^2+1)q^2(\vec{k}+\vec{p}+\vec{q})^2 } \; .
\label{3.9}
\ee
We decouple the $(\vec{k},\vec{p},\vec{q})$ integrals by writing
\ba
{1\over (\vec{k}+\vec{p}+\vec{q})^2}&=&\int d^3w
{\delta^3(\vec{k}+\vec{p}+\vec{q}+\vec{w})\over w^2}\nn\\
&=&\int {d^3w \over w^2} \int  {d^3r \over (2\pi)^3}
e^{i \vec{r} \cdot (\vec{k}+\vec{p}+\vec{q}+\vec{w})}\,\,.
\label{3.10}
\ea
Inserting eq.(\ref{3.10}) into eq.(\ref{3.9}), the
$(\vec{w},\vec{k},\vec{p},\vec{q})$
integrals become trivial, giving
\ba
\mbox{eq.(\ref{3.9})} &= &\int d^3r \left(1-e^{-r} \over 4\pi r \right)^2
\left( 1\over 4 \pi r
\right)^2 \label{3.11}\\
&=&{1\over(4\pi)^3} \int^\infty_0 {dr\over r^2} (1-e^{-r})^2 \nn\\
&=&{ (-1)^2\over (4\pi)^3} {\cal{J}}_2(\alpha\to 0) \nn\\
&=&{1\over(4\pi)^3} [2 \ln 2]\,\,. \label{3.12}
\ea
We have defined a function
\ba
{\cal{J}}_n(\alpha)&\equiv& \int^\infty_0 {dr\over r^2} (e^{-r}-1)^n
e^{-\alpha r} \nn\\
&=&\sum^{n}_{k=0} (-1)^k {n\choose k} (n+\alpha-k) \ln (n+\alpha-k) \,\,,
\label{3.13}
\ea
which will appear repeatedly. The use of eq.(\ref{3.10})
in (\ref{3.9}) is equivalent to evaluating
the momentum integrals (\ref{3.9}) in coordinate space (\ref{3.11})
 (see \cite{AZ}). In (\ref{3.11}) one
 recognises $1\over4\pi r$ as the coordinate space Coulomb
propagator and $e^{-r}\over 4\pi r$ as the coordinate space screened-
Coulomb propagator. Adding eqs.(\ref{3.8}) and (\ref{3.12}) gives

\be
I_2= {g^4 m T^3 \over (4\pi)^3} 2\ln 2 +
{g^4 m^2 T^2 \over 2^8\ \pi^4\ \ep} +  O(g^6)\,.
\label{3.14}
\ee
Finally, $I_3$ and $I_4$ are both finite and their evaluation is
analogous to the steps leading from  eq.(\ref{3.9}) to (\ref{3.12}) and
utilises eq.(\ref{3.13}). We find
\ba
I_3&=& -g^4 m T^3 {\cal{J}}_{3}(\alpha \to 0) =
{g^4 m T^3 \over (4\pi)^3} [3\ln3 - 6\ln2] + O(\ep g^5) \; ,\label{3.15}\\
I_4&=& g^4 m T^3 {\cal{J}}_{4}(\alpha \to 0) =
{g^4 m T^3 \over (4\pi)^3} [20\ln2 - 12\ln 3] + O(\ep g^5) . \label{3.16}
\ea
Combining eqs.(\ref{2.6},\ref{3.2},\ref{3.6},\ref{3.14}-\ref{3.16})
we get
\ba
\mu^{2 \ep} {\cal{P}}_h &=&
{g^4 T^4 \over 3^3\ 2^{12}\ \pi^2}\left({6\over\ep} +18\ln {\mu^2\over 4\pi
T^2}
-12{\zeta^{\prime}(-3)\over\zeta(-3)} +
48{\zeta^{\prime}(-1)\over\zeta(-1)} -18\gamma +{182\over5}\right) \nn \\
&&-{g^5 T^4 \over 3\ \sqrt{6}\ 2^{11} \pi^3}\left({1\over\ep} +6\ln {\mu\over
m}
+8 -3\gamma -4\ln{2} + 3 \ln{\pi} \right) \nn \\
&&+{g^6 T^4 \over 3\ 2^{14}\ \pi^4} {1\over\ep} \, + \, O (g^6) \, .
\label{3.17}
\ea
The full $g^5$ contribution is then obtained by adding this to the
value of diagrams (a-g), some of which also contain
$g^5$ pieces,  given in eqns.(\ref{2.2}-\ref{2.5}).
The result
is displayed in the following section.

\section{Summary of results}
\label{SR}

The sum of diagrams gives
\ba
{\cal{P}}&=&{\pi^2 T^4\ov 9}\bb {1\ov 10} - {1\ov8}
\left(g\ov4\pi \right)^2 +{1\ov\sqrt{6}}\left(g\ov 4\pi \right)^3 \nn\\
& & \nn\\
& & - \left( g\ov 4\pi \right)^4 \left[ - {3\ov 16}\ln{\mu^2\ov 4\pi T^2}
 + {1\ov 4} {\zeta^{\prime}(-3)\ov \zeta(-3)} -{1\ov 2}{\zeta^{\prime}(-1)\ov
\zeta(-1)} + {\gamma\ov 16} + {59\ov 120} \right] \nn\\
& & \nn\\
& &+ \left( g\ov 4\pi \right)^5 \sqrt{3\ov 2}\ \left[
-{3\ov4}\ln{ \mu^2\ov 4 \pi T^2} + {\zeta^{\prime}(-1)\ov \zeta(-1)}
+{\gamma-5\ov 4}  + \ln{g^2\ov 24 \pi^2} \right]\eb + O(g^6)\,\,. \label{4.1}
\ea

As required, all divergences, including the spurious $g^6/\ep$
terms, have cancelled. The renormalisation scale $\mu$ appears
explicitly in the $\ln \left(\mu\over T\right)$ terms and also
implicitly in the coupling constant, $g\,=\,g(\mu)$ . One can
eliminate the $\ln \left(\mu\over T\right)$ terms
by re-expressing
the pressure in terms of the
temperature dependent coupling $g(T)$ given by

\be
g^2(T)=g^2(\mu)\left[ 1+{3 g^2(\mu) \over (4\pi)^2} \ln
{T\over\mu}\right] + O(g^6) \; .
\label{4.2}
\ee
Then
\ba
{\cal{P}}&=&{\pi^2 T^4\ov 9} \bb {1\ov 10} - {1\ov8}
\left( g(T)\ov 4\pi \right)^2 +{1\ov \sqrt{6}} \left(g(T)\ov 4\pi
\right)^3\nn\\
& & \nn\\
& & -\left( g(T)\ov 4\pi \right)^4 \left[ {3\ov 16}\ln 4\pi
+{1\ov 4}{\zeta^{\prime}(-3)\ov \zeta(-3)}-{1\ov 2}{\zeta^{\prime}(-1)\ov
\zeta(-1)} +{\gamma\ov 16} +{59\ov 120}\right] \nn\\
& & \nn\\
& &+ \left( g(T)\ov 4\pi \right)^5 \sqrt{3\ov 2}\ \left[
{3\ov4}\ln { 4 \pi } +{\zeta^{\prime}(-1)\ov \zeta(-1)}
+{\gamma-5\ov 4} + \ln{g^2(T) \ov 24\pi^2}\right]\eb + O(g^6)\,\,.
\label{4.3}
\ea
A simpler but perhaps less instructive way to obtain eq.(\ref{4.3}) is
 to choose $\mu=T$ in eq.(\ref{4.1}).

\section{Conclusion}
\label{conclusion}
Using the minimally rearranged Lagrangian (\ref{1.3}) together
with an $\epsilon$-independent thermal mass $m^2={g^2 T^2\over 24}$,
we have verified previous \cite{FST,AZ} fourth order results for the
pressure of massless $g^2 \phi^4$ theory
and then extended the calculations to fifth order. Our final
result is given by eqs.(\ref{4.1}) and (\ref{4.3}). At the fifth order
a coupling constant logarithm appears for the first time in the
pressure of $g^2 \phi^4$ theory.
For QCD such coupling constant logarithms appear
already at fourth order, but they do not occur  in the
pressure of QED (at zero chemical potential) because there are
no self-interactions of photons (the only soft fields in imaginary
time) and the conclusion follows by power counting \cite{P1}.

It is natural to contemplate next a fifth order calculation in QCD.
Based on our experience with $g^2\phi^4$ theory and QED we expect
such a calculation to be technically simpler than the corresponding
fourth order (three-loop) calculation: Nontrivial three-loop diagrams
which are IR finite in the bare theory and computationally
 difficult (e.g. $I_0$ in $g^2\phi^4$
theory), contribute at subleading order ($g^5$) when at least one of
bare propagators is replaced by a zero-mode dressed propagator
(c.f. $I_1$ in eq.(\ref{i1})) so that the sum-integral over the
dressed momentum line collapses to a
three-dimensional  UV finite integral, leaving only two
overlapping frequency sums  at most. Since  frequency sums
are the main complication in these calculations, this reduction saves effort.
In practice further simplification
 has been observed : for the fifth order QED calculation,
summing over gauge-invariant sets of diagrams
results in the cancellation \cite{P1} of the terms with two
overlapping frequency sums; for the scalar calculation in this paper
the terms with two-overlapping frequency sums were found to contribute
at higher order ($g^6$)
(see the evaluation of $I_1$). That is, both the QED
and scalar fifth order calculations turned out to be easier than expected.
This bonus might prevail for QCD.

As noted by Linde \cite{L} many years ago, the perturbative
evaluation of the pressure in QCD breaks down at order $g^6$ because
of the absence of magnetic screening at lowest order. Braaten
\cite{B} has recently proposed a solution whereby one can obtain the
coefficient of the $g^6$ contribution as a functional integral in a
dimensionally reduced effective theory obtained by integrating out
the hard fields in the original QCD path integral.
Braaten has also described how his effective Lagrangian may be used
to obtain the lower order $g^5$ term and it would be  interesting to
compare the result of that approach with one using a shifted
Lagrangian  {\it \`{a} la } eq.(\ref{1.3}). \\
\begin{center}
**********
\end{center}
\vspace{0.5cm}

{\bf {\large Figure Caption\\}}
Fig.1(a-h) : Diagrams which contribute to fifth order. The propagators
are given by eq.(4), the cross represents the thermal counterterm of
eq.(3) while the blob in diagram(d) is the ultraviolet counterterm.


\end{document}